\documentclass[preprint,12pt]{elsarticle}

\usepackage{graphicx}
\usepackage{epstopdf}
\usepackage{bm}
\usepackage{mathrsfs}
\usepackage{amssymb}
\usepackage{lineno}

\journal{Nuclear Physics A}

\begin{document}

\begin{frontmatter}

\title{Nuclear matter calculations with the phenomenological three-nucleon interaction}

\author{H. Moeini\corref{mycorrespondingauthor}}
\cortext[mycorrespondingauthor]{Corresponding author}
\address{Department of Physics, School of Science,
Shiraz University,\\ Shiraz 71946-84795, Iran}
\ead{h.moeini@shirazu.ac.ir}

\author{G. H. Bordbar}
\address{Department of Physics, School of Science,
Shiraz University,\\ Shiraz 71946-84795, Iran (Permanent address),\\
 and\\
Department of Physics and Astronomy, University of Waterloo, 200 University Avenue West, Waterloo, Ontario, N2L 3G1, Canada} \ead{ghbordbar@shirazu.ac.ir}
 
\begin{abstract}
Employing the concept of three-body radial distribution function and using the two-body correlation functions,
calculated based on the lowest order constrained variational method, we investigated the effect of the three-body force (TBF) on the nuclear matter properties, for Argonne and Urbana $\it{v_{14}}$ potentials. As such, the results for nuclear matter density, incompressibility, energy per nucleon, and symmetry energy are presented at the saturation point. The inclusion of a phenomenological TBF resulted in closer values of the saturation density, incompressibility, and symmetry energy to the empirical ones for the symmetric nuclear matter. This is especially the case for the Urbana $\it{v_{14}}$ potential. In addition, an empirically-verified parabolic approximation of the interaction energy was utilized to perform an approximate study of the nuclear matter with neutron excess. Hence, at densities higher than about 0.3~fm$^{-3}$ and for proton-to-neutron density ratios close to the symmetric nuclear matter, the inclusion of TBF resulted in an extra attraction for the Argonne as compared to the Urbana $\it{v_{14}}$ potential.
\end{abstract}


\end{frontmatter}

\section{Introduction}
\label{sec:intro}
Classes of theoretical models which allow for predicting nuclear matter properties mainly include phenomenological density functionals, ab-initio many-body calculations, and effective field theories (EFTs). In a picture matching more closely our present understanding of the structure of nucleons, one cannot rule out the existence of many-body effects. In other words, as nucleons are made up of quarks and gluons with internal degrees of freedom, it is natural to consider the possibility of density-dependent interactions among three and more nucleons. Since the use of modern $NN$ potentials in reproducing the three- and four-nucleon data results in $\chi^2$ values much larger than those describing two-nucleon systems, the concept of three-nucleon forces has proven vital in calculating the properties of symmetric nuclear matter at equilibrium. Furthermore, the effect of three-nucleon forces in high-density neutron matter and nucleon matter equation of state (EOS) is expected to become large and hence influential in a realistic description of dense stars, supernovae, and high-energy heavy-ion collisions~\cite{Kievsky2010,Lovato2012,WiringaFiks1988}. The maximum-mass properties of neutron stars can be affected considerably by the TBF influence at very high densities, as the maximum mass depends strongly on the equation of state~\cite{BordbarHayati2006}. In this regard, neutron stars can serve as astrophysical laboratories for evaluating theories of dense nuclear matter. This is especially interesting in the light of recent developments in neutron-star physics and the discovery of massive neutron stars of about 1.97~\cite{Demorest2010}, 2.01~\cite{Antoniadis2013}, and 2.3~\cite{Linares2018} solar masses. Such masses are more than most of the neutron star masses observed so far $(1.2~–~1.6~M_{\odot})$ in binary neutron star systems~\cite{Ozel2012,Foucart2020}, which is challenging many of the present models of the EOS at high densities.

Relativistic effects could introduce characteristic qualities for the EOS, in comparison with the non-relativistic approaches~\cite{Brockmann1990}. For instance, the Dirac-Brueckner-Hartree-Fock (DBHF) approach which is based on the realistic NN interactions, allows to deal with short-range NN correlations that are missing in mean-field theories. Its success in reproducing the saturation point without any tuning of the coupling constants is mainly due to the inclusion of density-dependent in-medium NN scattering amplitudes, short range correlations as well as relativistic many-body effects~\cite{Katayama2013}. Such qualities can prove useful and important for investigating dense nucleonic systems like neutron stars whose core densities could reach several times that of the nuclear matter. In such high density baryonic matter, the nucleon chemical potential increases rapidly with density which might instantiate degrees of freedom heavier than nucleons -- such as strange matter -- so that the existence of hyperons (Y) might be energetically favored~\cite{Vidana2018}, leading to the softening of the EOS~\cite{Inoue2019}. Such effect is estimated, in many of the theoretical models involving hyperons~\cite{Serot1986}, to lower the maximum mass of the neutron stars to values below what recent observations~\cite{Demorest2010,Antoniadis2013,Linares2018} revealed (hyperon puzzle) and below the values predicted by scenarios in which only nucleons and leptons are assumed to exist in the core. The puzzle might indicate that the creation of hyperons in neutron stars should be considerably suppressed~\cite{Miyatsu2013} or that the hyperon-hyperon repulsive interaction would make the EOS stiffer~\cite{Char2015}, so that the resulting EOS would not be so softened after all, and the hyperon puzzle would no longer exist. In either case, it suggests the need to address the effect of density-dependent in-medium correlations on the EOS of the core through methods such as DBHF and Density-Dependent Relativistic Hadron (DDRH) field theory with the inclusion of hyperons. Furthermore, in the absence of exact experimental measurements of the hyperon-nucleon (YN), let alone hyperon-hyperon (YY) potentials, Lattice-QCD calculations prove to be an emerging vital tool to constrain such interactions~\cite{Drischler2020}. Thus, it is important to combine such techniques with many-body approaches to determine the EOS more precisely~\cite{Beane2012}. 

Microscopic and phenomenological TBF formalisms have been exploited to explain properties such as the light nuclei binding energies, saturation point of nuclear matter, and the spin dynamics in the nucleon-deuteron scattering~\cite{Lombardo2004}. The phenomenological Urbana-Argonne (UA) TBF models are particularly used in variational calculations for finite nuclei as well as nuclear matter, and are composed of a two-pion-exchange potential and a phenomenological repulsive term~\cite{Carlson1983,Pudliner1995}. The assumption herein has been that the attractive two-pion-exchange term and the repulsive term would dominate in $^3$H/$^4$He systems and in high-density nuclear matter, respectively~\cite{LagarisPandharipande1981_2}. This can be understood, for example, since realistic two-body models tend to underbind $^3$H by an order of 1~MeV~\cite{GibsonMcKellar1988}. As an indirect indication, there have been also studies arguing that the existence of neutron stars with masses of about two times the Earth’s mass would rule out nuclear models with no TBF assumptions~\cite{Akmal1998}.

The lowest order constrained variational method (LOCV) has been developed for $v_8$~\cite{LagarisPandharipande1980}, $v_{12}$ and Urbana $v_{14}$ (UV14)~\cite{LagarisPandharipande1981}, Argonne $v_{14}$ (AV14)~\cite{WiringaSmith1984}, and Argonne $v_{18}$ (AV18)~\cite{WiringaStokes1995} potentials with similar outcomes to other variational methods that include contributions from many-body clusters~\cite{BordbarModarres1997,BordbarModarres1998}.
The LOCV formalism has particularly a couple of vantage points over many-body techniques such as the variational~\cite{LagarisPandharipande1981_2,WiringaFiks1988}, Brueckner-Bethe (BB)~\cite{DayWiringa1985}, and Brueckner-Hartree-Fock~\cite{BaldoBombaci1988} methods. Having no free parameters and introducing a normalization constraint in its formalism, makes LOCV a fully self-consistent variational method while keeping the higher-order cluster terms comparatively small and as such rendering the method computationally less expensive than other many-body techniques. The correlation functions are calculated within LOCV by solving the Euler-Lagrange equation, which is obtained by functional minimization of the energy. Besides, the Lagrange multiplier corresponds to the normalization constant of the wave function. These have made LOCV a parameter-free variational method in which the calculations are relatively simplified with respect to other many-body techniques.
Within the LOCV framework and exploiting various two-body potentials such as the Reid, $\Delta$-Reid, AV14, UV14, AV18, Reid $v_{8}$, and Reid $v_{12}$, we have extensively studied the EOS and incompressibility of asymmetric nuclear matter~\cite{BordbarModarres1998,ModarresBordbar1998}. These studies comprised also calculating the effect of the three-nucleon interactions (TNI) through including the three-body term of the energy functional cluster expansion~\cite{BordbarModarres1997}. As such, our investigations of zero- and finite-temperature nuclear matter have covered various areas including bulk properties of nuclear matter in the light of charge-independence breaking in nucleon-nucleon interaction~\cite{Bordbar2003}, magnetized neutron matter~\cite{BordbarRezaei2013} and magnetic susceptibility of nuclear matter~\cite{BigdeliBordbarRezaei2009,RezaeiBordbar2017}, spin polarized neutron and nuclear matter~\cite{BordbarBigdeli2007,BordbarBigdeli2008,BordbarRezaeiMontakhab2011,RezaeiBigdeliBordbar2015} and asymmetric nuclear matter~\cite{BordbarBigdeli2008_2nd,BordbarBigdeli2007_2nd,BigdeliBordbarPoostforush2010} in the context of neutron star research subsuming saturation properties, dynamical stability~\cite{HendiBordbar2016,BordbarHendi2016,RezaeiBordbar2016,HendiBordbar2017}, expansion, and contraction of magnetized and cold neutron stars~\cite{EslamPanahBordbar2017,EslamPanahYazdizadeh2019}.

The LOCV formalism with the sole inclusion of the two-body force, like other such calculations, cannot reproduce the empirical characteristics of symmetric nuclear matter at saturation point
\cite{BordbarModarres1997,BordbarModarres1998}. 
Therefore, as nuclei are, in general, under-bound when fits of only NN potentials to the scattering data are considered, the many-body Hamiltonian in this formalism is constructed as the sum of the total kinetic energy of nucleons plus all nucleon-pair potentials plus three-nucleon potentials, assuming the latter term to approximate the main contribution from many-nucleon forces. Within the LOCV approach, the contribution of the three-nucleon potential can be accounted for either microscopically or phenomenologically. Throughout this article, the use of the term TNI would refer to either microscopic calculations or those phenomenological treatments of the TBF which do not completely follow the construction of the UA models. In the case of entries of Table~\ref{tab:table1} which correspond to “variational” method of calculation (row \#5 and \#7), TNI is approximated by adding density-dependent repulsive and attractive terms to the UV14 potential. The repulsive term is taken to be the product of the intermediate-range part of the two-body interaction and an exponential factor whose negative argument is assumed to be proportional to the density. Introducing such exponential term would not only reduce effectively the intermediate-range attraction of the two-nucleon force but it also allows approximating higher-order many-body forces as the terms of a series with alternating signs. The attractive two-pion-exchange term is treated phenomenologically, assuming it to be proportional to the product of squared value of the density and an exponential function with a negative argument proportional to the density. Such exponential term would guarantee the negligibility of the attractive term at higher densities. In the case of LOCV entry of Table~\ref{tab:table1} (row \#8), TNI refers to the calculation of the three-body term of the energy functional cluster expansion. 

The TBF effects in the UA models are usually modeled by adjusting the strength of the long-range two-pion exchange and intermediate-range repulsive terms to render the best fit to the saturation point of the nuclear matter and binding energies of light nuclei. However, three-body forces are in principle different for different nuclei due to, among others, the density-dependent in-medium interactions which makes it difficult to simultaneously fit all these quantities. Hence, we followed the procedure of Baldo {\it{et al.}}~\cite{Baldo1999} and constrained the two strength parameters to reproduce only the nuclear matter saturation point. This is performed, utilizing the concept of three-body radial distribution function as well as introducing nucleon-nucleon correlations in the calculations. As such, we derived the EOS for the symmetric as well as asymmetric nuclear matter with the help of an empirical parabolic approximation of the interaction energy. It should be mentioned that we have performed three-body calculations using the concept of three-body distribution function. Hence, in order to put into perspective this approach toward the phenomenological TBF calculation within the LOCV framework, we studied the AV14 and UV14 potentials in the presence of the obtained TBF effect, hereafter referred to as TBF$_{UA}$. Details of other UA-type phenomenological treatments of the TBF that are referred to as UVII and UIX are addressed by the corresponding references in Table~\ref{tab:table1}.

This paper continues as follows: We first give a brief review of the two- and three-body potentials in the UA models, for the nuclear matter at zero temperature. Next, we provide an overview of the calculation of energy contributions from the two-body potential and TBF$_{UA}$, using the correlation functions derived within the LOCV formalism. Finally, we discuss the results and summarize the conclusions.

\section{Inter-nucleon interactions}
\label{sec1}
The many-body nuclear theory assumes that the effects of all subnucleonic degrees of freedom related to mesons and nucleonic resonances can be absorbed into nuclear forces, which makes the following a good approximation, i.e. ignoring the four-body terms etc., of the low-energy Hamiltonian below pion production threshold~\cite{NATOASIseries1997}:
\begin{equation}
H=\sum_{{i\leq A}}\frac{-\hbar^2}{2m}\nabla^2_i+\sum_{{i< j\leq A}}V_{ij}+\sum_{{i< j< k\leq A}}V_{ijk}.
\label{Eq:Hamiltonian}
\end{equation}
in which $V_{ij}$ and  $V_{ijk}$ are the two-body and three-body potentials, respectively.

The two-body potential is constrained by $NN$ scattering data and deuteron properties.
The two-nucleon interaction, in the UA $\it{v_{14}}$ models, is considered as follows:
\begin{eqnarray}
V(12)=\sum_{p=1}^{14}v^{(p)}(r_{12})O^{(p)}_{12}
\label{v12}
\end{eqnarray}
where, the fourteen operators $O^{(p)}_{12}$ are given as follows~\cite{LagarisPandharipande1981}:
\begin{eqnarray}
O^{(1,2,\dots,14)}_{12}=\textbf{1},{\boldsymbol\sigma}_1\cdot{\boldsymbol\sigma}_2,{\boldsymbol\tau}_1\cdot{\boldsymbol\tau}_2,
({\boldsymbol\sigma}_1\cdot{\boldsymbol\sigma}_2)({\boldsymbol\tau}_1\cdot{\boldsymbol\tau}_2),\textbf{S}_{12},\textbf{S}_{12}
({\boldsymbol\tau}_1\cdot{\boldsymbol\tau}_2),\nonumber\\
(\textbf{L}\cdot\textbf{S}),(\textbf{L}\cdot\textbf{S})({\boldsymbol\tau}_1\cdot{\boldsymbol\tau}_2),
\textbf{L}^2,\textbf{L}^2({\boldsymbol\sigma}_1\cdot{\boldsymbol\sigma}_2),\textbf{L}^2({\boldsymbol\tau}_1\cdot{\boldsymbol\tau}_2),\nonumber\\
\textbf{L}^2({\boldsymbol\sigma}_1\cdot{\boldsymbol\sigma}_2)({\boldsymbol\tau}_1\cdot{\boldsymbol\tau}_2),
{(\textbf{L}\cdot\textbf{S})}^2,{(\textbf{L}\cdot\textbf{S})}^2({\boldsymbol\tau}_1\cdot{\boldsymbol\tau}_2)~~~~~~~~~~~
\end{eqnarray}
in which
\begin{eqnarray}
\textbf{S}_{ij}=3({\boldsymbol\sigma}_i\cdot\hat{\textbf{r}}_{ij})({\boldsymbol\sigma}_j\cdot\hat{\textbf{r}}_{ij})-{\boldsymbol\sigma}_i\cdot{\boldsymbol\sigma}_j.
\end{eqnarray}

The three-body potential in Eq.~(\ref{Eq:Hamiltonian}) includes two attractive and repulsive parts. In the UA models, the attractive long-range part of three-body potential, $V_{ijk}=V_{ijk}^{2\pi}+V_{ijk}^R$, corresponds to a two-pion exchange interaction that was first investigated by Fujita and Miyazawa~\cite{FujitaMiyazawa1957,GibsonMcKellar1988,WiringaFiks1988}:
\begin{eqnarray}
 V_{ijk}^{2\pi}=A_{2\pi}\sum_{cyc}\Big(\{X_{ij}^{\pi},X_{ik}^{\pi}\}\{{\boldsymbol\tau}_i\cdot{\boldsymbol\tau}_j,
{\boldsymbol\tau}_i\cdot{\boldsymbol\tau}_k\}+ \frac{1}{4}[X_{ij}^{\pi},X_{ik}^{\pi}][{\boldsymbol\tau}_i\cdot{\boldsymbol\tau}_j,{\boldsymbol\tau}_i\cdot{\boldsymbol\tau}_k]\Big)
\label{Eq:A2pi}
\end{eqnarray}
in which
\begin{eqnarray}
X_{ij}^{\pi}&=&Y_{\pi}(r_{ij}){\boldsymbol\sigma}_i\cdot{\boldsymbol\sigma}_j+T_{\pi}(r_{ij})\textbf{S}_{ij},
\label{Eq:Xij}
\end{eqnarray}
where,
\begin{eqnarray}
Y_{\pi}(r)&=&\frac{e^{-m_{\pi}r}}{m_{\pi}r}\big(1-e^{-cr^2}\big),\nonumber\\
T_{\pi}(r)&=&\Big(1+\frac{3}{m_{\pi}r}+\frac{3}{m_{\pi}^2r^2}\Big)Y_{\pi}(r)\big(1-e^{-cr^2}\big).\nonumber
\end{eqnarray}
The intermediate-range repulsive part of $V_{ijk}$ is, in the UA models, phenomenologically considered as follows~\cite{LagarisPandharipande1981_2}:
\begin{equation}
V_{ijk}^R=U\sum_{cyc}T^2_{\pi}(r_{ij})T^2_{\pi}(r_{ik})
\label{Eq:U0}
\end{equation}
The constants $A_{2\pi}$ and $U$ are adjusted to reproduce the empirical saturation point of nuclear matter.
The cut-off $(1-e^{-cr^2})$ is based on the phenomenological $NN$ force models and represents the modification of one-pion-exchange potential at small $r$ by the finite size of the nucleons and pions~\cite{GibsonMcKellar1988,NATOASIseries1997}.
The cut-off constant is set $c=2.1$~fm$^{-2}$, for the $v_{14}$ potentials~\cite{Pudliner1995,Schiavilla1986,WiringaStokes1995}. In this work, $m_{\pi}=\frac{1}{3}(m_{\pi^0}+2m_{\pi^{\pm}})c/\hbar$ is taken as the average of the pion masses, for which the above constants are obtained to be $A_{2\pi}^{(AV14)}=-0.0329$, $U^{(AV14)}=0.0064$, $A_{2\pi}^{(UV14)}=-0.0331$, and $U^{(UV14)}=0.0045$~MeV.

\section{Energy calculations for nuclear matter}
\label{sec:EnergyCalculations}

Using the variational approach, the expectation value of energy can be written as follows,
\begin{eqnarray}
E&=&\frac{\langle\Psi|H|\Psi\rangle}{\langle\Psi|\Psi\rangle}\nonumber\\
&=& E_K + \bigg\langle\sum_{ij} V_{ij}\bigg\rangle + \bigg\langle\sum_{ijk} V_{ijk}\bigg\rangle,
\label{ExpectHamiltonian}
\end{eqnarray}
where $E_K$ is the kinetic energy contribution, and $\bigg\langle\sum_{ij} V_{ij}\bigg\rangle$ and $\bigg\langle\sum_{ijk} V_{ijk}\bigg\rangle$ are the contribution of the two- and three-body nucleon-nucleon potentials, respectively. At zero temperature, the kinetic energy contribution per nucleon is given by
\begin{equation}
\frac{E_K}{A}=\frac{3{\hbar^2}{k_{F}^2}}{10m},
\end{equation}
where the Fermi wave-number $k_F$ is related to the number density of protons ($\rho_p$) and neutrons ($\rho_n$) as well as the spin-isospin degeneracy $\nu$ as  $\rho=\rho_n+\rho_p=\nu k_F^3/6\pi^2$.
Our approaches for calculating the contribution of the two- and three-body nucleon-nucleon potentials are as follows.

\subsection{Contribution of two-body nucleon-nucleon potential}
\label{subsec:TwoBodyPotential}

Adopting a variational wave function of the form $\Psi=F\Phi$, in which $\Phi$ and $F(1\dots A)$ are, respectively, the ground-state wave function and a correlation operator for a total number of $A$ independent nucleons, each of mass $m$ occupying a total volume of $A/\rho$, one can approximate the cluster expansion of the (ground-state) energy. In accordance with the cluster expansion of the energy functional up to the two-body term, the following relation holds for the two-body $NN$ potential energy \cite{Clark1979}:
\begin{eqnarray}
\bigg\langle\sum_{ij} V_{ij}\bigg\rangle=\frac{1}{2}\sum_{{ij}}\langle ij|\mathscr{V}(12)|ij-ji\rangle
\label{ExpectVij}
\end{eqnarray}
in which
\begin{equation}
\mathscr{V}(12)\equiv \frac{-\hbar^2}{2m}\big[f(12),[\nabla_{12}^2,f(12)]\big]+f(12)V(12)f(12)
\label{V12}
\end{equation}
where $V(12)$ is the bare two-nucleon interaction defined in Eq.~(\ref{v12}) and the two-body correlation operators $f(12)$, which act on the spin, isospin, and relative position variables of particles 1 and 2, are considered as follows,
\begin{equation}
f(12)=\sum_{\alpha,p=1}^{3}f_{\alpha}^{(p)}(12)\mathscr{O}_{\alpha}^{(p)}(12),~\alpha\in\{J,L,S,T,T_z\}
\label{f12}
\end{equation}
For singlet and triplet uncoupled channels with $J=L$, $p$ is chosen 1 and $\mathscr{O}_{\alpha}^{(1)}(12)=\textbf{1}$. For triplet coupled channels with $J=L\pm1$, $p$ is set equal to 2 and 3 for which
\begin{equation}
\mathscr{O}_{\alpha}^{(2,3)}(12)=
(\boldsymbol{\frac{2}{3}}+\frac{1}{6}\textbf{S}_{12}),(\boldsymbol{\frac{1}{3}}-\frac{1}{6}\textbf{S}_{12})
\end{equation}
Here, $\textbf{S}_{12}$ is the tensor operator introduced in Eq.~(\ref{Eq:Xij}). The correlation functions $f_{\alpha}^{(i)}(r)$ can be calculated by imposing the following normalization condition \cite{Clark1979} on the two-body distribution function $g(\textbf{r}_1,\textbf{r}_2)$ and by minimizing $E_2$ with respect to $f_{\alpha}^{(i)}(r)$.
\begin{eqnarray}
\rho\int d^3r_{12}\Big(1-g(\textbf{r}_1,\textbf{r}_2)\Big)= \rho\int d^3r_{12}\Big(1-f^2(r_{12})g_{F}(r_{12})\Big)=1
\end{eqnarray}
in which, the two-body radial distribution function of the non-interacting Fermi-gas ground state is given by:
\begin{equation}
g_{F}(r_{12})=1-\frac{1}{\nu}l^2(k_Fr_{12})
\end{equation}
with
\begin{eqnarray}
l(k_Fr_{ij})=\frac{\nu}{N}\sum_{k<k_F}e^{i\textbf{k}\cdot(\textbf{r}_j-\textbf{r}_i)}
=3\Bigg(\frac{\sin(k_Fr_{ij})-k_Fr_{ij}\cos(k_Fr_{ij})}{{(k_Fr_{ij})}^3}\Bigg)
\end{eqnarray}

\subsection{Contribution of three-body nucleon-nucleon potential}
\label{subsec:ThreeBodyForce}

Using the three-body radial distribution function, we can calculate the expectation value of the three-nucleon potential, Eq.~(\ref{ExpectHamiltonian}), as follows:
\begin{eqnarray}
\langle V_{3}\rangle \equiv \bigg\langle\sum_{ijk} V_{ijk}\bigg\rangle
=\frac{\langle\Psi|\sum_{ijk} V_{ijk}|\Psi\rangle}{\langle\Psi|\Psi\rangle}~~~~~~~~~~~~~~~~~~~~~~~~~~~~~~~~~~~~~~~~~~~~~~~~~~~~ \nonumber\\
=\sum_{\sigma_1 \cdots \sigma_N}\sum_{\tau_1 \cdots \tau_N} \int\cdots\int d\textbf{r}_1 \cdots d\textbf{r}_N \Psi^*(\textbf{r}_1 \sigma_1 \tau_1,\cdots,\textbf{r}_N \sigma_N \tau_N)~~~~~~~~~~~~~~~~~~~~ \nonumber\\
\times\sum_{i<j<k} V(\textbf{r}_i,\textbf{r}_j,\textbf{r}_k) \Psi(\textbf{r}_1 \sigma_1 \tau_1,\cdots,\textbf{r}_N \sigma_N \tau_N)~~~~~~~~~~~~ \nonumber\\
=\frac{N(N-1)(N-2)}{3!}\sum_{\sigma_1 \cdots \sigma_N}\sum_{\tau_1 \cdots \tau_N} \int{\cdots}\int d\textbf{r}_1 \cdots d\textbf{r}_N ~~~~~~~~~~~~~~~~~~~~~~~~~~~ \nonumber\\
\times\Psi^*(\textbf{r}_1 \sigma_1 \tau_1,\cdots,\textbf{r}_N \sigma_N \tau_N) V(\textbf{r}_1,\textbf{r}_2,\textbf{r}_3) \Psi(\textbf{r}_1 \sigma_1 \tau_1,\cdots,\textbf{r}_N \sigma_N \tau_N) \nonumber \\
\label{expValV3}
\end{eqnarray}
in which $V(\textbf{r}_1,\textbf{r}_2,\textbf{r}_3)$ is the effective three-body potential $V_{123}$, substituted here with the sum of the values of the two terms defined in Eqs.~(\ref{Eq:A2pi},\ref{Eq:U0}). We compute the interaction between the three particles as the summation of three terms, one of which accounting for the interaction between particle\#1 and the other two particles considered as a whole. The other terms of the summation are obtained by changing the three particle indices cyclically. In fact, the summation of the values of Eq.~(\ref{Eq:U0})
and operators (Eq.~(\ref{Eq:A2pi})) in the spin and isospin states give the value of the three-nucleon potential in spin and isospin states, namely $V(\textbf{r}_1,\textbf{r}_2,\textbf{r}_3)$ which substitutes in Eq.~(\ref{expValV3}). Now, defining the three-body density matrix as
\begin{eqnarray}
\Gamma(\textbf{r}_1 \sigma_1 \tau_1,\textbf{r}_2 \sigma_2 \tau_2,\textbf{r}_3 \sigma_3 \tau_3)=\frac{N(N-1)(N-2)}{3!}\times~~~~~~~~~~~~~~~~~~~~~~~~~~~~~~~~~~~~~~~~ \nonumber\\
\sum_{\sigma_4 \cdots \sigma_N}\sum_{\tau_4 \cdots \tau_N} \int \cdots \int d\textbf{r}_4 \cdots d\textbf{r}_N
\Psi^*(\textbf{r}_1 \sigma_1 \tau_1,\cdots,\textbf{r}_N \sigma_N \tau_N)  \Psi(\textbf{r}_1 \sigma_1 \tau_1,\cdots,\textbf{r}_N \sigma_N \tau_N) \nonumber\\
\label{GammaFunction}
\end{eqnarray}
results in:
\begin{eqnarray}
\langle V_{3}\rangle = \sum_{\sigma_1 \sigma_2 \sigma_3}\sum_{\tau_1 \tau_2 \tau_3} \int\int\int d\textbf{r}_1 d\textbf{r}_2 d\textbf{r}_3 V(\textbf{r}_1,\textbf{r}_2,\textbf{r}_3) \Gamma(\textbf{r}_1 \sigma_1 \tau_1,\textbf{r}_2 \sigma_2 \tau_2,\textbf{r}_3 \sigma_3 \tau_3) \nonumber\\
\end{eqnarray}
Hence, employing the definition of the three-body radial distribution function $g(\textbf{r}_1,\textbf{r}_2,\textbf{r}_3)$ as
\begin{equation}
\rho^3 g(\textbf{r}_1,\textbf{r}_2,\textbf{r}_3)=3! \sum_{\sigma_1 \sigma_2 \sigma_3}\sum_{\tau_1 \tau_2 \tau_3} \Gamma(\textbf{r}_1 \sigma_1 \tau_1,\textbf{r}_2 \sigma_2 \tau_2,\textbf{r}_3 \sigma_3 \tau_3)
\label{rho3g}
\end{equation}
reformulates the expectation value of the three-nucleon potential as follows:
\begin{equation}
\langle V_{3}\rangle
=\frac{\rho^3}{6}\int d\textbf{r}_1 d\textbf{r}_2 d\textbf{r}_3 V(\textbf{r}_1,\textbf{r}_2,\textbf{r}_3)g(\textbf{r}_1,\textbf{r}_2,\textbf{r}_3)
\end{equation}
The three-body radial distribution function can be calculated in a self-consistent way as~\cite{Clark1979}
\begin{eqnarray}
g(\textbf{r}_1,\textbf{r}_2,\textbf{r}_3)=f^2(r_{12})f^2(r_{23})f^2(r_{13})g_{F}(\textbf{r}_1,\textbf{r}_2,\textbf{r}_3)
\label{gFunction}
\end{eqnarray}
Here, $g_{F}(\textbf{r}_1,\textbf{r}_2,\textbf{r}_3)$ is the three-body radial distribution function of the non-interacting Fermi-gas ground state, given as
\begin{eqnarray}
g_{F}(\textbf{r}_1,\textbf{r}_2,\textbf{r}_3)
=1-\frac{1}{\nu}\big[l^2(k_Fr_{12})+l^2(k_Fr_{23})+l^2(k_Fr_{13})\big]\nonumber\\
+\frac{2}{\nu^2}l(k_Fr_{12})l(k_Fr_{23})l(k_Fr_{13})~~~~~~~~~~~~~
\end{eqnarray}
and the effect of the inter-nucleon interactions is taken into account through inter-particle correlation functions $f(ij)$, which have been extracted within the LOCV framework~\cite{BordbarModarres1998}.

\section{Results}
\label{sec:results}

\subsection{Binding energy}
\label{results:1}

\begin{figure*}[h]
\includegraphics[width=0.95\textwidth]{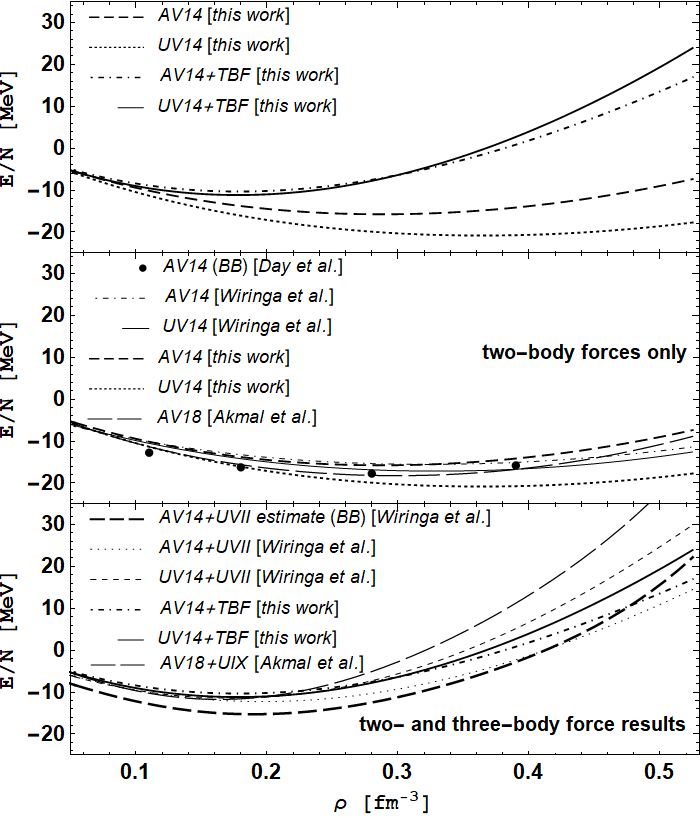}
\caption{\label{fig:bindingE_nuclearMatter} Top: our results for the binding energy per nucleon for symmetric nuclear matter, as a function of nucleon density. Middle: comparison between the two-body calculations of this work, Wiringa {\it{et al.}}~\cite{WiringaFiks1988}, and Akmal {\it{et al.}}~\cite{Akmal1998}. The latter results were extracted from the supplied tables in the article, by employing polynomial fits to the AV18 and AV18+UIX data. The results labeled as ``BB'' are from Day {\it{et al.}}~\cite{DayWiringa1985}. Bottom: same as the middle panel, with the inclusion of the three-body contribution.}
\end{figure*}

Fig.~\ref{fig:bindingE_nuclearMatter} compares various calculations of the binding energy per nucleon as a function of density, for symmetric nuclear matter. The results are provided for the UV14, AV14, and AV18 potentials, with and without the three-body contribution, which are calculated based on the phenomenological UVII and UIX models. Our two-body UV14 and AV14 results for the nuclear matter saturation energy per nucleon are about -20.8 and -15.7~MeV, respectively. These calculations with the inclusion of TBF$_{UA}$ yield values of about -11.2 and -10.3~MeV, respectively. The corresponding saturation densities obtained from these calculations for the UV14 (UV14+TBF$_{UA}$) and AV14 (AV14+TBF$_{UA}$) models are about 0.364 (0.178) and 0.288 (0.178)~fm$^{-3}$, respectively. Overall, the results of different potentials and works in Fig.~\ref{fig:bindingE_nuclearMatter} indicate that the effect of adding TBF$_{UA}$ to the two-body potential increases the core stiffness of the effective potential. It is interesting to note that the symmetric nuclear matter results of Wiringa {\it{et al.}}~\cite{WiringaFiks1988} predict a lower stiffness for the AV14 than for the UV14 potential, as a result of including the three-body effect UVII. This seems not to be the case, in absence of the three-body contribution. Our results show the same behaviour, although, contrary to the results of Wiringa {\it{et al.}}, the difference between our results for the two potentials seems to become less pronounced when TBF$_{UA}$ is included and more pronounced in the absence of TBF$_{UA}$. Incidentally, the results of Akmal {\it{et al.}}~\cite{Akmal1998} for the AV18 potential are overall, in the absence of the three-body effect UIX, more than (less than) our UV14 (AV14) results. Adding the three-body contribution seems, for densities above about $0.2~\rm{fm}^{-3}$, to have made the AV18 effective potential of Akmal {\it{et al.}} stiffer than other UA $\it{v_{14}}$ effective potentials that include the three-body contribution. As a result, at lower densities, the AV18+UIX potential appears to be slightly more bound than other potentials that include the three-body contribution.

\subsection{Symmetry energy}
\label{results:2}

\begin{figure*}[h]
\includegraphics[width=0.95\textwidth]{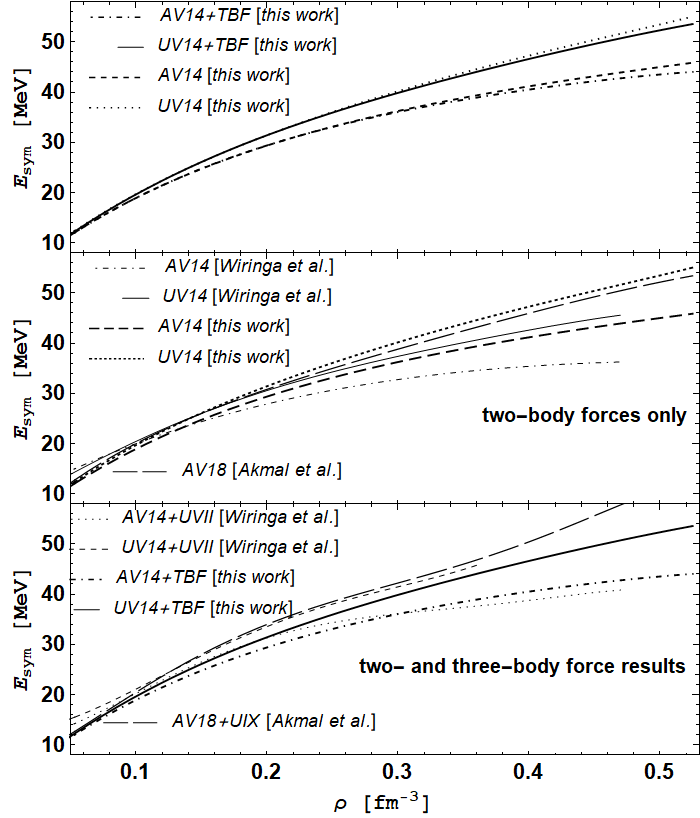}
\caption{\label{fig:symmetryE} Top: our results for the symmetry energy as a function of density for different potentials. Middle: comparison between the two-body calculations of this work, Wiringa {\it{et al.}}~\cite{WiringaFiks1988}, and Akmal {\it{et al.}}~\cite{Akmal1998}. Bottom: same as the middle panel, with the inclusion of the three-body contribution.}
\end{figure*}

Assuming a quadratic dependence of the EOS on asymmetry, the symmetry energy is equal to the difference between the EOS of the pure neutron matter and symmetric nuclear matter. Fig.~\ref{fig:symmetryE} shows the results for the symmetry energy obtained based on our energy calculations. Our two-body (and two-body with the inclusion of TBF$_{UA}$) results of the symmetry energy at the saturation density for the UV14 and AV14 models are about 44.8 (29.2) and 35.5 (27.6)~MeV, respectively.

\begin{figure*}[h]
\includegraphics[width=0.95\textwidth]{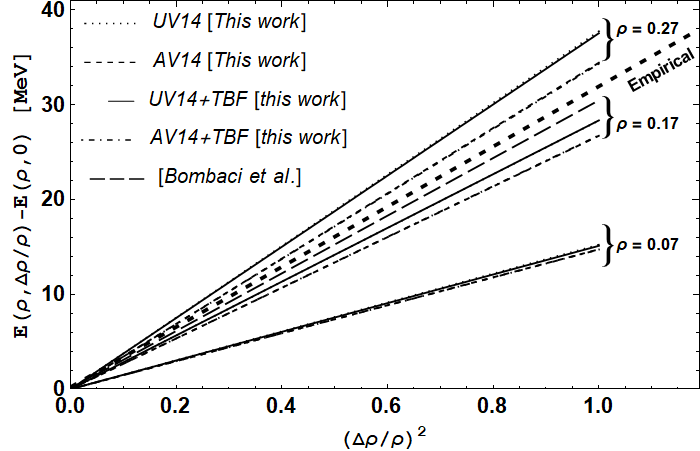}
\caption{\label{fig:quadraticAssymetryEnergy} Quadratic dependence of asymmetric energy for different potentials and nucleon densities. The data are from this work and Bombaci {\it{et al.}}~\cite{BombaciLombardo1991}. The two-body data of this work coincide almost with the corresponding data when the TBF$_{UA}$ effect is included.}
\end{figure*}

\begin{figure*}[h]
\includegraphics[width=0.82\textwidth]{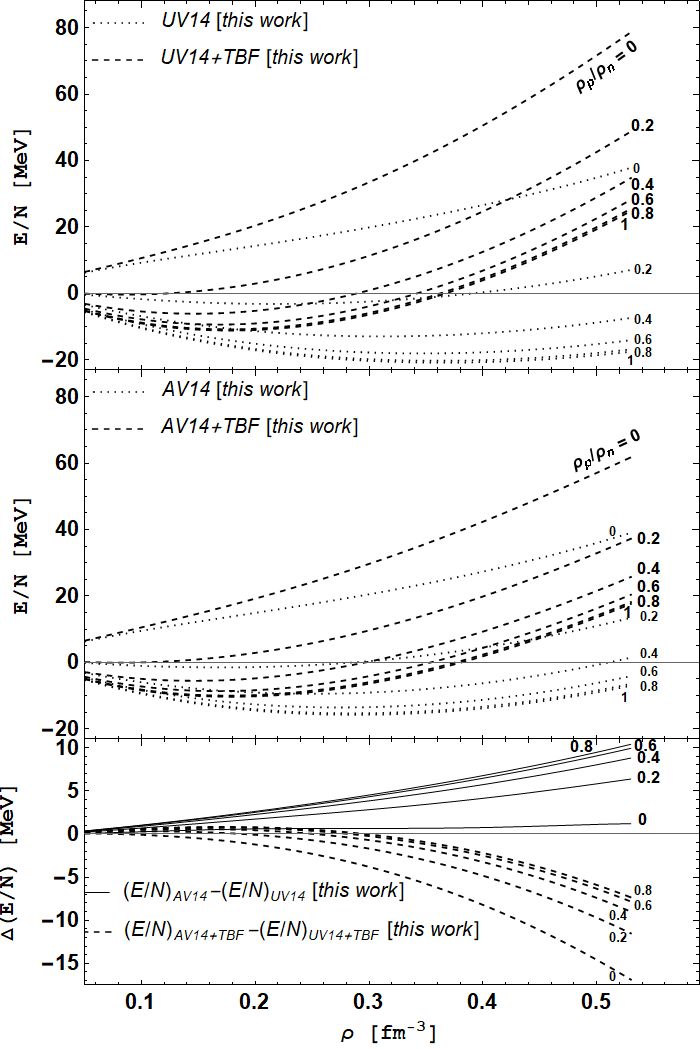}
\caption{\label{fig:2And23BodyBindingE_asymmetricMatter}Top panels: approximate LOCV calculations of the two-body binding energy, with and without TBF$_{UA}$ contribution, for the asymmetric nuclear matter and UA $\it{v_{14}}$ potentials, as a function of density. Bottom panel: difference between the approximate binding energies of AV14 and UV14 potentials with (broken lines) and without (solid lines) the inclusion of TBF$_{UA}$. Numbers next to the curves indicate the proton to neutron ratios of 0, 0.2, 0.4, 0.6, 0.8, 1, which correspond to $\Delta\rho/\rho$ values of 1, 0.67, 0.43, 0.25, 0.11, 0, respectively.}
\end{figure*}

Asymmetric nuclear matter is interesting in many respects, particularly in the context of astrophysics involving the collapse of massive stars. Employing the $v_{14}$ two-body potentials and a phenomenological three-body interaction, Lagaris {\it{et al.}}~\cite{LagarisPandharipande1981_3} have reported on extending the variational calculations of symmetric nuclear matter to treat the asymmetric nuclear matter through a parabolic approximation as follows:
\begin{equation}
E(\rho,\Delta\rho/\rho)\approx E(\rho,0)+E_{sym}(\rho)(\Delta\rho/\rho)^2
\label{parabolicApprox}
\end{equation}
where $\Delta\rho/\rho=(\rho_{n}-\rho_{p})/(\rho_{n}+\rho_{p})$ indicates the level of nuclear matter asymmetry and is assumed to vary between 0 and 1, corresponding to the symmetric nuclear matter and pure neutron matter, respectively. Eq.~(\ref{parabolicApprox}) is experimentally confirmed at least for small values of $\Delta\rho/\rho$, although some theoretical works have shown that it is fulfilled at high values of $\Delta\rho/\rho$ as well~\cite{BombaciLombardo1991}. Assuming that the parabolic form of the binding energy per nucleon is valid for $0\leq\Delta\rho/\rho\leq1$, Fig.~\ref{fig:quadraticAssymetryEnergy} shows the resulting quadratic dependence of the asymmetry energy $E(\rho,\Delta\rho/\rho)-E(\rho,0)$ on $\Delta\rho/\rho$, for different potentials and nucleon densities. Indeed, the quadratic relation of our data in Fig.~\ref{fig:quadraticAssymetryEnergy} are obtained using Eq.~(\ref{parabolicApprox}) and the symmetry energy results of Fig.~\ref{fig:symmetryE}, allowing to approximate the EOS for the asymmetric nuclear matter based only on the EOS of pure neutron and symmetric nuclear matter. Our results for the UV14 potential at the saturation density are closer to the empirical values than those for the AV14 potential. 
Fig.~\ref{fig:2And23BodyBindingE_asymmetricMatter} shows the parabolic approximation results for the binding energy of the asymmetric nuclear matter of different proton to neutron ratios, based on the LOCV calculations for the UA $\it{v_{14}}$ potentials as well as the phenomenological TBF$_{UA}$ calculations. The results in the bottom panel show that, at a given density, as the ratio of protons to neutrons $\rho_p/\rho_n$ increases, the UV14 potential yields ever more attraction than AV14. 
This seems not to be the case when TBF$_{UA}$ is included; at higher densities for all $\rho_p/\rho_n$ or at all densities for lower values of $\rho_p/\rho_n$, the UV14+TBF$_{UA}$ potential appears to yield less attraction than AV14+TBF$_{UA}$. In addition, as the ratio of protons to neutrons increases, the difference in the attraction of the two potentials appear to decrease.
Furthermore, at a given $\rho_p/\rho_n$, the difference in the binding energy of UV14 and AV14 increases with density linearly.
The corresponding results, when TBF$_{UA}$ is included, appear to change with density rather exponentially.
Comparing with the exact calculations for the case of two-body force inclusion~\cite{BordbarModarres1998}, an approximate curve in Fig.~\ref{fig:2And23BodyBindingE_asymmetricMatter} would approach more closely to the corresponding exact solution, as the value of $\Delta\rho/\rho$ gets smaller or $\rho_p/\rho_n$ becomes larger. Such behavior should be expected as well after the inclusion of TBF$_{UA}$. Hence, based on the latter conclusion that the results of the bottom panel of Fig.~\ref{fig:2And23BodyBindingE_asymmetricMatter} should be closer to the exact calculations for closer values of $\rho_p/\rho_n$ to unity, we conclude that at densities higher than about 0.3~fm$^{-3}$ and for proton-to-neutron density ratios close to the symmetric nuclear matter, the inclusion of TBF$_{UA}$ would result in an extra attraction for the Argonne as compared to the Urbana $\it{v_{14}}$ potential. A comparison between the approximate calculations with the inclusion of only two-body force and the corresponding exact results of Bordbar {\it{et al.}}~\cite{BordbarModarres1998} reveals that, for a given $\rho_p/\rho_n$, larger deviations of the approximate solutions from the exact ones happen at larger densities and that there is a maximum of less than 1~MeV difference between their predicted binding energies, for $\rho_p/\rho_n\gtrsim 0.6$. This difference between the exact and approximate solutions is less than about 0.1~MeV, for $\rho_p/\rho_n\gtrsim0.8$.

\subsection{Pressure and incompressibility}
\label{results:3}

\begin{figure*}[h]
\includegraphics[width=0.95\textwidth]{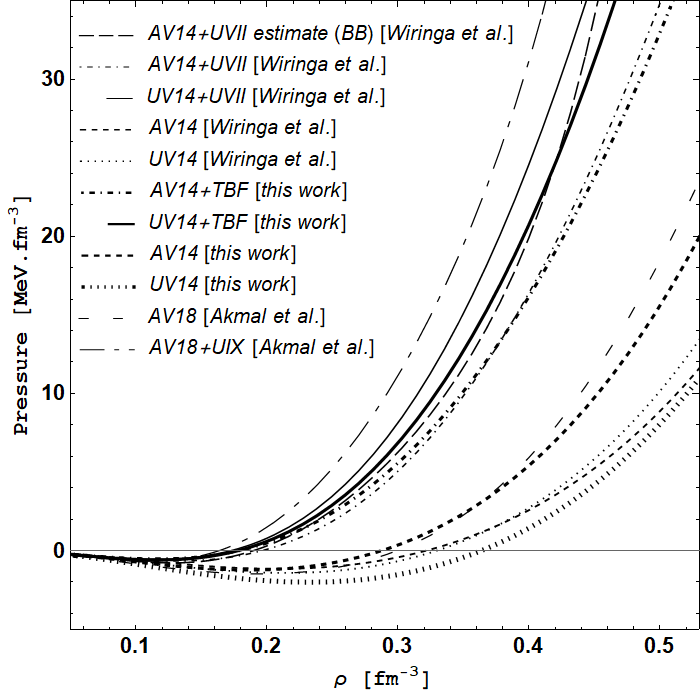}
\caption{\label{fig:symmetricNuclearMatterPressure}Pressure of the symmetric nuclear matter for different potentials, as a function of density. The data of Wiringa {\it{et al.}}~\cite{WiringaFiks1988} and Akmal {\it{et al.}}~\cite{Akmal1998} are extracted from their binding energy results for the symmetric nuclear matter.}
\end{figure*}

\begin{figure*}[h]
\includegraphics[width=0.9\textwidth]{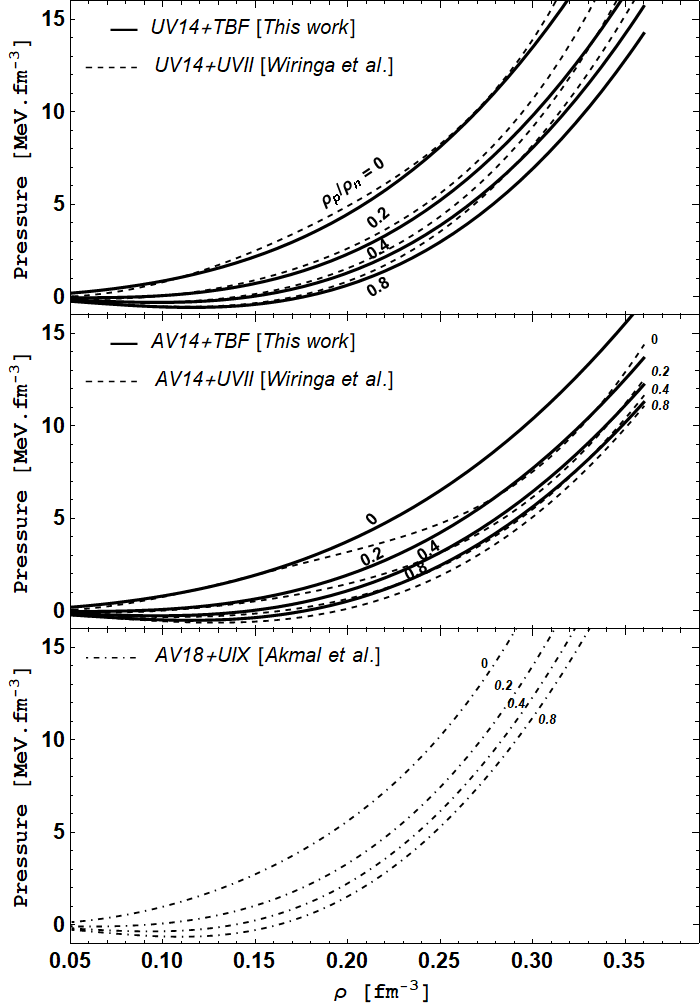}
\caption{\label{fig:P23}Approximate LOCV two-body calculations with the inclusion of TBF$_{UA}$, for the pressure of the asymmetric nuclear matter with different proton to neutron ratios, as a function of density. The data of Wiringa {\it{et al.}}~\cite{WiringaFiks1988} and Akmal {\it{et al.}}~\cite{Akmal1998} are extracted from their binding energy results for the pure neutron and symmetric nuclear matter.}
\end{figure*}

The pressure of asymmetric nuclear matter, at a proton and neutron density of $\rho_{p}$ and $\rho_{n}$, is given as follows:
\begin{equation}
P={\rho}^2\frac{\partial{E(\rho_{p},\rho_{n})}}{\partial{\rho}}
\end{equation}
in which $\rho=\rho_{p}+\rho_{n}$.
Fig.~\ref{fig:symmetricNuclearMatterPressure} puts in perspective the pressure results of various potentials and methods, for the symmetric nuclear matter. The results of different potentials and works indicate that the effect of adding the TBF contribution results in more pressure in the nuclear medium. In other words, the effect of including TBF is in the direction of adding to the core stiffness of the effective potential, which is in accordance with the results of Fig.~\ref{fig:bindingE_nuclearMatter}. Fig.~\ref{fig:P23} shows our parabolic approximation results for the pressure of the asymmetric nuclear matter of different proton to neutron ratios, based on the LOCV calculations for the UA $\it{v_{14}}$ potentials with the TBF$_{UA}$ contribution. According to the exact calculations of Bordbar {\it{et al.}} with two-body force inclusion~\cite{BordbarModarres1998}, the difference between the approximate and exact solutions grows with $\Delta\rho/\rho$. Hence, in accordance with the discussion regarding Fig.~\ref{fig:2And23BodyBindingE_asymmetricMatter}, the results in Fig. \ref{fig:P23} should be cautiously considered for $\rho_p/\rho_n\lesssim0.6$.

The incompressibility of infinite nuclear matter $K_{\infty}$, denoted here as $K_0$ for the incompressibility at the saturation density $\rho_0$, relates to the curvature of the energy per particle of the symmetric nuclear matter at $\rho_0$:
\begin{equation}
K_0=9\rho_0^2 \frac{\partial^2(E/A)}{\partial\rho^2}\Bigg|_{\rho_0}
\end{equation}
The interest in $K_0$ is associated with the interest in the physics of supernovae evolution and neutron stars. Using the measured value of the isoscalar giant monopole resonance in $^{208}$Pb, the non-relativistic results of Skyrme energy functionals and Gogny functionals have been shown to be consistent and point to a value of $230-240$~MeV~\cite{Colo2004,Colo2004_2,Blaizot1995}, which is not in agreement with the relativistic mean field predictions of $250-270$~MeV~\cite{Ma2001,Niksic2002}, derived from $^{208}$Pb and $^{144}$Sm iso-scalar monopole data. The model dependence of the incompressibility amounts to a discrepancy of $10-20\%$ in the $K_0$ values obtained within different models, though there are suggestions that such difference between the predictions of the relativistic and non-relativistic models is significantly less than $20\%$~\cite{Agrawal2003}. Recent calculations of the centroid of the measured giant monopole resonance in $^{208}$Pb and $^{120}$Sn yields a $K_0$ value of $230\pm 40$~MeV~\cite{Khan2012}. There has been also indications that the non-relativistic and relativistic models can be reconciled, which places the value of $K_0$ at $240\pm 10$~MeV~\cite{Piekarewicz2010}. Our calculations, with two-body force inclusion using the UV14 and AV14 potentials, result a saturation incompressibility of about 302 and 240~MeV, respectively. The results with the inclusion of TBF$_{UA}$ are obtained to be about 193 and 167~MeV for UV14 and AV14, respectively. In accordance with the pressure results of Fig.~\ref{fig:symmetricNuclearMatterPressure}, one can expect that the inclusion of TBF$_{UA}$ would increase the overall incompressibility of $9\rho^2 \frac{\partial^2(E/A)}{\partial\rho^2}$ as a function of density.

\section{Discussion and conclusions}
\label{sec:conclusions}

Employing the UA $v_{14}$ potentials within the LOCV formalism, we have performed calculations for the EOS of nuclear matter at zero temperature, taking into account a TBF effect in accordance with the phenomenological UA models. 
Investigating the previously well-studied potentials AV14 and UV14 allowed us to focus primarily on the influence of the newly constructed TBF$_{UA}$, in comparison with alternative approaches towards these potentials that have been taken to estimate the TBF contribution. In spite of the known limitations regarding the chosen potentials in the present work, investigating the implications of the proposed TBF$_{UA}$ suggests utilizing it for further studies of realistic potentials like AV18 and other modern nucleon-nucleon models which fit the Nijmegen database, built based on various nn and np scattering data points below 350 MeV, with a reduced $\chi^2$ of about unity.

For different potentials and calculational methods, Table~\ref{tab:table1} shows the properties of symmetric nuclear matter including saturation energy per nucleon, saturation density, saturation incompressibility, and symmetry energy at the saturation density. The results labeled by row \#28 are based on the local chiral potential of Piarulli {\it{et al.}}~\cite{Piarulli2016} at N3LO of ChPT, which includes $\Delta$ isobar excitations and employs the Brueckner-Bethe-Goldstone (BBG) many-body theory within the Brueckner-Hartree-Fock (BHF) approximation. The results of row \#23 and \#24 are based on the DBHF approach, using the Bonn potentials A and B and employing polynomial parameterizations for the EOS of symmetric nuclear matter in terms of nuclear density.
Requiring the potential to reproduce the nuclear-matter saturation point through adjusting the strength parameters $A_{2\pi}$ and U of the phenomenological TBF$_{UA}$, we have derived the EOS of the symmetric nuclear matter by employing the concept of three-body radial distribution function in association with the two-body correlations that were extracted within the LOCV framework. The correlation functions $f(r_{ij})$ approach unity, where there is no interaction between the two particles as $r_{ij}$ approaches infinity. Clearly, $f(r_{ij}\rightarrow0)$ approaches zero, where the potential between the two particles approaches infinity. Eqs.~\ref{GammaFunction} and \ref{rho3g} introduce the three-body density matrix $\Gamma$ and highlight its relation with the three-body radial distribution function $g$. In this respect, the three-body distribution function of Eq.~(\ref{gFunction}) has allowed us to perform non-relativistic three-body calculations without treating TBF as an effective two-body interaction scheme as discussed in other studies \cite{Grange1989,Lejeune1986,Goudarzi2018}. Hence, nuclear matter parameters such as the symmetry energy, pressure, and incompressibility were calculated to gain an understanding of their behavior as a result of including TBF$_{UA}$. In principle, the strength parameters of a phenomenological TBF depend on the adopted theoretical method and the two-body interaction, and they should be adjusted to give a best fit to nuclear matter properties within the adopted theoretical framework -- here, the LOCV method. 

In comparison with the two-body results, the inclusion of TBF$_{UA}$ seems to have shifted the parameter values in Table~\ref{tab:table1} appreciably. As such, the incompressibility values have decreased significantly, with a closer value to the experimental results for UV14+TBF$_{UA}$, as compared with AV14+TBF$_{UA}$. Incidentally, the inclusion of TBF$_{UA}$ has resulted in saturation energies farther to the empirical value, as compared to the calculations with the sole inclusion of the two-body force. This is because the TBF$_{UA}$ influence is in the direction of increasing the core stiffness of the effective potential and hence results in less attraction that is reflected in the saturation-energy results as well. On the other hand, inclusion of the TBF$_{UA}$ term has moved the symmetry energy much closer to the empirical value. This seems to be more pronounced, in the case of the UV14 potential. An analysis of the pressure and incompressibility results for the symmetric nuclear matter shows that the TBF$_{UA}$ effect, as compared to lack thereof, increases the pressure and overall incompressibility of the nuclear matter at a given density.

In the specific case of AV14 (rows \#11 and \#12), the inclusion of TBF has resulted in farther binding energy, incompressibility, and symmetry energy values to the empirical ones. Having an ultimately positive TBF contribution to the EOS of infinite nuclear-matter means that constraining the strength parameters would not necessarily result in reproducing both of the empirical saturation energy and density. In other words, as the AV14 binding energy is already close to the empirical value, any additional positive contribution on top of the two-body EOS (due to the TBF effect) that is capable of shifting the two-body saturation density to considerably lower values around the empirical one would consequently result in less-bound nuclear matter. This can be understood, given the shape of the two-body EOS in Fig.~\ref{fig:bindingE_nuclearMatter} and the fact that the TBF effect amounts to positive contributions on top of the overall two-body EOS. Same behavior can be seen also in the data of Wiringa {\it{et al.}}~\cite{WiringaFiks1988} and Akmal {\it{et al.}}~\cite{Akmal1998}, provided in Fig.~\ref{fig:bindingE_nuclearMatter} and Table~\ref{tab:table1}. Nevertheless, in the case of UV14 (rows \#9 and \#10), only the saturation energy is estimated (-11.15~MeV) to be as distant to the empirical value (-16~MeV) as the corresponding two-body result (-20.82~MeV) is. It should be mentioned that the objective of this study has not been to state the general relevance of TBF as a potential; though, there are calculations either based on TNI (for instance, for UV14 in Table~\ref{tab:table1}, rows \#5 and \#8; and for AV18, row \#20) or based on phenomenological UA models (for instance and to some extent, for AV18 in Table~\ref{tab:table1}, rows \#18 and \#15; and for UV14, row \#10), which do not entirely rule out the relevance of three-body correlations for the infinite nuclear matter. Rather, our objective has been to introduce and check the implications of a form of TBF derived utilizing three-body radial distribution function associated with two-body correlation functions.

The sizable deviations -- with respect to the empirical results -- of the binding energy due to inclusion of TBF$_{UA}$ does not seem to be the characteristics of only AV14 potential. This can also be seen, though less pronounced, in the UV14 and AV18 results of Table~\ref{tab:table1}, rows \#4 (UV14+UVII), \#15 (AV18+UIX), and \#10 (UV14+TBF$_{UA}$). As to the deviation of binding energies upon the inclusion of three-nucleon forces, it seems that, overall, the inclusion of TNI with respect to lack thereof (for instance, compare the results of row \#5 and row \#3, or row \#8 and row \#9, or row \#20 and row \#19) shows less deviations as compared with the phenomenological UA models (for instance, compare the results of row \#4 and row \#3, or row \#10 and row \#9, or row \#15 and row \#14). It is, however, also notable that the results of the rows \#17 and \#18 (TBFa \& BHF method) indicate about the same amount of deviation as the results of row \#19 and \#20 (TNI \& BHF method) do. This could suggest that the considerable deviation of the results of row \#15 from the two-body calculations in row \#14 has to do with the method of calculation rather than the use of phenomenological UIX. In fact, even the results of row \#16 (AV18+$\delta v$+UIX), which differ from those of row \#15 by including a relativistic boost correction $\delta v$ to the two-nucleon interaction, appear to do not much better (except for the symmetry energy results) than the results of AV18+UIX (row \#15), when put in comparison with empirical results. The small perturbative correction $\delta v$ relates to an improvement of the variational wave function, which happens when one calculates the correlation functions separately in every $J,~S,~l$ channel~\cite{Akmal1997}. The boost interaction is regarded as part of the two-nucleon interaction and, hence, the models indicated in rows \#15 and \#16 are both regarded realistic~\cite{Akmal1998}. However, in the case of the results of row \#18, the deviation of the binding energy from the corresponding two-nucleon value in row \#17 is less than the deviations of the binding energy of row \#15 or even its relativistic version (row \#16) with respect to row \#14. This is interesting, considering the fact that the results of row \#18 are based on the (non-relativistic) BHF method, using the AV18 and a similar UA-type TBF potential as the one in row \#15 or \#16. 

The saturation-point predictions in Table~\ref{tab:table1} allow for comparing the LOCV results with other calculations on the market. As an example, the AV18 results of row \#13 and 
\#14 which are based on LOCV and VCS, respectively, appear to be closely comparable. Also, the results of row \#19 which are based on the BHF method can be compared with the LOCV results of row \#13. Furthermore, employing the same potential and including the three-body effects calculated microscopically, the LOCV and variational results of row \#8 and row \#5 are also closely comparable. As an example with a different potential, one can compare the AV14 results of row \#11 and 
\#1, which are based on the LOCV and variational methods, respectively. Of course, these conclusions on the proximity of different model results are made aside from the fact that none of these calculations have reported their uncertainties to allow for a more robust comparison. Hence, ignoring the influence of the uncertainties, the saturation-point predictions of the latter models appear to be in good agreement. This is the case, however, with the exception of the incompressibility, which is clearly a sensitive parameter in the sense that the comparison of the two values should be done on an equal footing, for which one would need to know the analysis details of both calculations such as theoretical errors and numerical uncertainties as a function of density. In this work, we did an investigation of the binning effect (number of EOS points calculated as a function of density) on the incompressibility values. As such, we compared the results for two different bin sizes, namely $\Delta\rho=0.005$ and 0.001~fm$^{-3}$ -- the latter of which being meant to represent fine binning. Thence, evaluating the effect of the errors of the fit parameters on the EOS allowed for calculating the standard errors of the incompressibility. For instance, for the AV14 potential, a $4^{th}$-order polynomial fit with $\Delta\rho=0.005$~fm$^{-3}$ would yield a $\chi^2_{R}=0.000354$, based on which the boundary values 240.2175 and 241.3458 are obtained around the centroid value 240.7816~MeV. Repeating the same procedure for $\Delta\rho=0.001$~fm$^{-3}$ yielded a $\chi^2_{R}=0.000286$, a centroid value of 240.2998, and boundary values of 239.8832 and 240.7166~MeV. Hence, based on a $4^{th}$-order polynomial fit to the EOS, we expect incompressibility standard errors of 0.56 and 0.42, corresponding to bin sizes $\Delta\rho=0.005$ and 0.001~fm$^{-3}$, respectively.

With regards to the model performance in presence of three-nucleon forces, the LOCV method has been already used for nuclear matter calculations taking into account the three-body term of the energy functional cluster expansion. The reported results of these calculations (row \#8) appear to be closely comparable with variational calculations of row \#5. It is notable that, in the presence as compared to the absence of TNI, the AV18 calculations of row \#20 has resulted both saturation energy and density to shift in the right direction. Incidentally, by including a phenomenological TBF in the form of UA models, none of the variational calculations in Table~\ref{tab:table1} seem to have done the justice in estimating both of the saturation energy and density closer to the empirical values. This includes also the AV18 results in rows \#14 and \#15. Thus, it is not conclusive from these calculations that sizable deviations of the results from empirical ones roots essentially in the type of potential (be it AV14 or UV14) but perhaps in a combination of the calculation method and the approach taken to calculate the TBF contribution. For instance, looking at the results in rows \#8 and \#9, it seems that the microscopic calculations of TBF within the LOCV framework have resulted in closer predictions of both saturation energy and density to the empirical values.  

In comparing various bulk properties of Table~\ref{tab:table1}, it should be noted that conclusions can be heavily influenced due to the meer fact that these quantities are obtained at the derived saturation densities within a specific method of calculation. Having said that, we notice that introducing TBF has consequential impacts (even up to 50\%) on the saturation properties including the density. Various microscopic studies have observed such effect~\cite{Zhou2004,Li2008,Li2008_2}, and indicated the more important effect of TBF on the isoscalar properties such as incompressibility than on the properties that are associated with the density dependence of the symmetry energy~\cite{Vidana2009}. Incidentally, these microscopic BHF calculations reveal, over the whole density range, consistent symmetry-energy results with some representative Skyrme forces; while such consistency with relativistic models shows up until around the saturation densities. Since many of the Skyrme forces are constructed to describe nucleonic systems in the vicinity of the saturation density, different Skyrme forces yield similar nuclear-matter EOS but different neutron-matter results. In fact, many of these and other relativistic effective models agree well in their predictions of the density, symmetry, and binding energy at saturation point, but disagree in other quantities like the one which portray the isospin dependence of the incompressibility~\cite{Piekarewicz2009}.

\section{Acknowledgements}
We wish to thank the Shiraz University Research Council.


\begin{table*}
\caption{\label{tab:table1}
Properties of symmetric nuclear matter including saturation energy per nucleon ($E_0/A$), density ($\rho_0$), incompressibility and symmetry energy at the saturation density ($K(\rho_0)$ and $E_{sym}(\rho_0)$), for different Hamiltonians and many-body techniques. For more details see the text. The incompressibility uncertainties in rows \#9 to \#12 are standard errors, calculated from a $4^{th}$-order polynomial fit to the EOS with a bin size of $\Delta\rho=0.005$~fm$^{-3}$.}
\begin{center}
{
\begin{tabular}{|c|cccccc|}
\hline
row&Interaction(s) &Calculation &$E_0/A$&$\rho_0$&$K(\rho_0)$&$E_{sym}(\rho_0)$\\
\#& &method & (MeV) & ($\rm{fm}^{-3}$) & (MeV) & (MeV)\\
\hline
1&AV14 & Variational~\cite{WiringaFiks1988}& -15.6 & 0.319 & 205 & 33.41 \\
2&AV14+UVII & Variational~\cite{WiringaFiks1988} & -12.4 & 0.194 & 209 & 30.83 \\
3&UV14 & Variational~\cite{WiringaFiks1988} & -17.1 & 0.326 & 243 & 38.83 \\
4&UV14+UVII & Variational~\cite{WiringaFiks1988} & -11.5 & 0.175 & 202 & 30.80 \\
5&UV14+TNI & Variational~\cite{WiringaFiks1988} & -16.6 & 0.157 & 261 & \\
6&AV14 & BB~\cite{DayWiringa1985,WiringaFiks1988} & -17.8 & 0.280 & 247 & \\
7&UV14+TNI & CBF~\cite{Fantoni1983,WiringaFiks1988} & -18.3 & 0.163 & 269 & \\
8&UV14+TNI & LOCV~\cite{BordbarModarres1997} & -17.33 & 0.170 & 276 & \\
9&UV14 & LOCV$^{\rm{this~work}}$ & -20.82 & 0.3636 & $301.86^{+0.15}_{-0.14}$ & 44.83 \\
10&UV14+TBF$_{UA}$ & LOCV$^{\rm{this~work}}$ & -11.15 & 0.1777 & $192.51\pm 0.21$ & 29.18 \\
11&AV14 & LOCV$^{\rm{this~work}}$ & -15.75 & 0.2876 & $240.78\pm 0.56$ & 35.53 \\
12&AV14+TBF$_{UA}$ & LOCV$^{\rm{this~work}}$ & -10.31 & 0.1785 & $167.09\pm 0.18$ & 27.56 \\
13&AV18 & LOCV~\cite{BordbarModarres1998} & -18.46 & 0.31 & 302 & \\
14&AV18 & VCS~\cite{Akmal1998} & -18.22 & 0.2993 & 291.0 & 38.75\\
15&AV18+UIX & VCS~\cite{Akmal1998} & -11.72 & 0.1634 & 249.8 & 29.78\\
16&AV18+$\delta v$+UIX & VCS~\cite{Akmal1998} & -12.17 & 0.1777 & 235.0 & 32.13 \\
17&AV18 & BHF~\cite{Vidana2009} & -17.3 & 0.24 & 213.6 & 35.8 \\
18&AV18+TBFa & BHF~\cite{Vidana2009} & -15.23 & 0.187 & 195.5 & 34.3 \\
19&AV18 & BHF~\cite{Li2006} & -17.3 & 0.259 & & 29.9\\
20&AV18+TNI & BHF~\cite{Zuo2002} & -15.08 & 0.198 & 207 & \\
21&Nijmegen-II & BHF~\cite{Li2006} & -19.4 & 0.326 & & 29.5\\
22&Reid-93 & BHF~\cite{Li2006} & -19.8 & 0.328 & & 30.0\\
23&CD-Bonn & BHF~\cite{Li2006} & -21.9 & 0.374 & & 31.1\\
24&Bonn A& DBHF~\cite{Katayama2013} & -16.62 & 0.181 & 233 & 34.8 \\
25&Bonn B& DBHF~\cite{Katayama2013} & -15.04 & 0.163 & 190 & 31.2 \\
26&$\rm{N^3LO}$ NN & $\chi$EFT~\cite{Bogner2009} & $\approx$-15.5 & $\approx$0.165 & $190-240$ & \\
27& $\rm{N^2LO}$ NN & $\chi$EFT~\cite{Lacour2011} & -16 & 0.17 & $240-250$ & \\
28& $\rm{N^3LO}$ NN+3N & $\chi$EFT~\cite{Kruger2013} & $-(14.1-21)$ & 0.17 & &$28.9-34.9$ \\
29&Nature &  & -16$\pm$1 & 0.17 & $230\pm40$~\cite{Khan2012} & $32\pm1$~\cite{Baldo2016}\\
\hline
\end{tabular}
}
\end{center}
\end{table*}

\end{document}